\documentclass{jpp}
\usepackage{graphicx}
\usepackage{epstopdf, epsfig}
\usepackage{hyperref}
\usepackage{amsmath,amsfonts,amssymb}

\usepackage{subcaption}

\def\U{{\boldsymbol U}}
\def\B{{\boldsymbol B}}
\def\emf{{\boldsymbol{E}_M}}
\def\J{{\boldsymbol J}}
\def\R{{\boldsymbol R}}
\def\V{{\boldsymbol V}}
\def\W{{\boldsymbol W}}
\def\Om{{\boldsymbol{\Omega}}}
\def\e{{\boldsymbol e}}
\def\u{{\boldsymbol u}}
\def\b{{\boldsymbol b}}

\def\n{{\boldsymbol n}}
\def\t{{\boldsymbol t}}

\def\S{{\boldsymbol S}}
\def\M{{\boldsymbol M}}

\def\d{{\rm d}}
\def\curlyC{{\mathcal C}}
\def\curlyS{{\mathcal S}}
\def\cur{{\boldsymbol j}}
\def\Re{{\rm Re}}
\def\Rm{{\rm Rm}}

\shorttitle{On turbulent magnetic reconnection}
\shortauthor{S. Stanish, D. MacTaggart}

\title{On turbulent magnetic reconnection: fast and slow mean steady-states}

\author{S. Stanish\aff{1}\corresp{\email{sage.stanish@glasgow.ac.uk}}, D. MacTaggart\aff{1}}

\affiliation{\aff{1} University of Glasgow, School of Mathematics and Statistics, Glasgow, Scotland}

\begin{document}

\maketitle

\begin{abstract}
We investigate a model of turbulent magnetic reconnection introduced by  Higashimori, Yokoi and Hoshino (Phys. Rev. Lett. 110, 255001) and show that the classic two-dimensional, steady-state Sweet-Parker and Petschek reconnection solutions are supported. We present evidence that these are the only two steady-state reconnection solutions, and we determine the criterion for their selection. Sweet-Parker reconnection occurs when there is no growth in turbulent energy, whereas Petschek reconnection occurs when the current density in the reconnecting current sheet is able to surpass a critical value, allowing for the growth of turbulent energy that creates the diffusion region. Further, we show that the Petschek solutions are self-similar, depending on the value of the turbulent time scale,  and produce a universal steady reconnection rate. The self-consistent development of Petschek reconnection through turbulence, within the model, is an example of fast and steady magnetic reconnection without an explicit need for the collisionless terms in an extended Ohm's law. 
\end{abstract}

\section{Introduction}

 Magnetic reconnection is a fundamental process of plasma physics. In astrophysical applications, it plays a crucial role in many important phenomena, including dynamo action, solar flares and the formation of coronal mass ejections (CMEs). In laminar and ideal magnetohydrodynamics (MHD), magnetic reconnection (hereafter, reconnection) is not possible - Alfv\'en's theorem constrains magnetic field lines to behave as material lines. To bypass this constraint, extra (non-ideal) terms need to be considered in Ohm's law. It is from this perspective, as a deviation from ideal MHD, that reconnection is often considered and has provided an excellent working definition for applications, particularly in relation to the solar atmosphere \citep{Pontin22}.

 Many of the physical processes that are based on reconnection, e.g. solar flares, are rapidly occurring events. Therefore, the reconnection associated with them is said to be \emph{fast}. This adjective needs to be understood relative to early work on reconnection, which was based on either linear instabilities or (approximate) steady-state solutions \citep{Priest07}. These theories result in reconnection rates that are too slow to describe fast phenomena such as flares.  As a result of this problem, a significant research trend developed that moved in the direction of seeking faster reconnection rates by considering the effects of extra physics (compared to resistive MHD) through an extended Ohm's law. The approach was exemplified by the Geophysical Environment Modelling (GEM) Magnetic Reconnection Challenge \citep{Birn01}  - a collection of works solving the same reconnection problem  (null point reconnection from the pinching of a current sheet) but with varying physical models. The main result of this study was that, for the given problem, the reconnection rates of all models are consistently faster than that of resistive MHD. This result, combined with those of other works \citep[e.g.][]{Biskamp86,Ma96,Uzdensky00,Malyshkin05} has led to a general consensus that  a key element for achieving fast reconnection is the inclusion of \emph{collisionless} effects.

 As well as the effects of microphysics, however, another fundamental property of astrophysical plasmas, which has a strong effect on reconnection, is \emph{turbulence}. Indeed, it has been argued that turbulence is of more importance than the extra terms of a generalized Ohm's law (i.e. the collisionless and resistive terms) as such terms are negligible compared to an inertial range electromotive force that is derived from ideal MHD \citep{Eyink15}. There are now several descriptions of turbulent magnetic  (fast) reconnection, including field line meandering (such as the theory of \cite{Lazarian99} and the simulations of \cite{2009ApJ...700...63K}), the dynamic formation of plasmoids (as described by \cite{uzdensky10} and simulated by \cite{2009MNRAS.399L.146L}) and enhanced transport modelling \citep[e.g.][]{Higashimori13}. We shall return to the latter of these shortly as this approach will form the basis for this work.

 Although reconnection is generally a three-dimensional (3D) process, historically, the categorization of slow and fast reconnection has been based on a (laminar) two-dimensional (2D) setup. The details of 2D reconnection solutions have been discussed at length elsewhere \citep[e.g.][]{Priest07}, so here we only provide some highlights that will be useful for our discussion later. Steady solutions of 2D reconnection follow either the \emph{Sweet-Parker} \citep{Parker57,Sweet58,Parker63} or the \emph{Petschek} \citep{Petschek64} configurations. Whilst the former solution is found for uniform resistive MHD, the latter is found when some extra physics or perturbation leads to the localized enhancement of the diffusion region, created thanks to what is often referred to as \emph{anomalous resistivity}. For turbulent reconnection, \cite{Higashimori13} employ a Reynolds-averaged and renormalized enhanced-transport turbulent MHD model. For 2D reconnection in a current sheet, they identify three solutions: a slow Sweet-Parker-like solution (denoted \emph{laminar reconnection}), a fast Petschek-like solution (denoted \emph{turbulent reconnection}) and a slower diffusive solution (denoted \emph{turbulent diffusion}). These three solutions were found (in order) by increasing the turbulent time scale (which is a free parameter).  In their model, it is the turbulent energy and cross-helicity that are responsible for  effectively providing anomalous resistivity and, thus, enabling the onset of Petschek reconnection. \cite{Widmer19} extended the model of \cite{Higashimori13} to include a model equation that determines the turbulent time scale, rather than choosing it as a free parameter. In their simulations, they resolved only the Petschek solution and concluded that the other two solutions are merely artefacts of the choice of turbulent time scale. However, based only on the simulations of \cite{Widmer19}, can the laminar and diffusive solutions found by \cite{Higashimori13} really be cast aside or do they form part of a more general overall picture of the solution space of this model? Further, can the steady-state reconnection solutions, that have formed the basis of much of the theory of 2D reconnection, be realized in this turbulent regime?

 The purpose of this work is to investigate the turbulent reconnection model (hereafter TRM) of \cite{Higashimori13}, in further detail, in order to develop a better understanding of how fast and slow reconnection develops within this framework and so address the points listed above. We perform two main tasks. First, we describe some general properties of the TRM in relation to field line topology, highlighting the role played by turbulence in changing the physics of reconnection with respect to laminar MHD. Secondly, we investigate the reconnection solutions described above and provide a detailed analysis for their selection. A corollary of this work is to demonstrate how effectively fast reconnection can be generated without the explicit need for the collisionless terms in a generalized Ohm's law.

 The layout of the paper is as follows. First, we introduce the TRM and summarize its main properties. Secondly, we describe some general properties of the TRM related to field line topology and reconnection. This is followed by a detailed study of simulations of 2D reconnection in a current sheet. In particular, we map out the solution space of the TRM for this problem and identify how individual solutions are selected.  We then provide some theoretical justification of the simulation results in relation to fast and steady reconnection. The paper ends with a summary and short discussion.

\section{The Turbulent Reconnection Model}

The TRM introduced in \cite{Higashimori13} consists of a set of MHD equations for mean fields together with one-point turbulent statistical quantities representing the turbulent fluctuations,  and provides a means of modelling turbulent reconnection in large-scale systems in which turbulence develops within the system as it evolves. The equations are derived using the two-scale direct-interaction approximation (TSDIA), for which detailed descriptions are presented in other works \citep[e.g.][]{yoshizawa13,Yokoi2020,Mizerski23}.  This closure scheme is for nonlinear inhomogeneous turbulence, for which it is assumed that fields can be split into mean and fluctuating parts. A multiple scales analysis is applied for large-scale inhomogeneities and closure is achieved by combining this with the direct-interaction approximation. We will make use of the notation for which a field $f$ is split into a mean part and a fluctuating part as $f=F+f'$, where upper-case letters represent mean quantities and lower-case letters with primes represent fluctuations of quantities.  Here, $F=\langle f\rangle$, where $\langle\cdot\rangle$ denotes an ensemble average.

In terms of the mean magnetic field $\B$ and the mean velocity field $\U$, the mean field incompressible MHD equations are

\begin{align}
	&\frac{\partial\U}{\partial t} + (\U\cdot\nabla)\U = -\nabla P + (\nabla\times\B)\times\B + \frac{1}{\Re}\nabla^2\U, \label{momentum} \\
	&\frac{\partial\B}{\partial t} =  \nabla\times(\U\times\B + \emf) + \frac{1}{\Rm}\nabla^2\B, \label{induction} \\
	&\nabla\cdot\B = \nabla\cdot\U = 0, \label{div_free}
\end{align}
where $P$ is the mean pressure, $\Re$ is the Reynolds number, $\Rm$ is the magnetic Reynolds number and $\emf$ is the electromotive force due to turbulent fluctuations. These equations are non-dimensional, based on scaling $\U$ with the (mean field) Alfvén speed $U_A$. The scales for length $L$ and the magnetic field $\B$ are based on the initial current sheet thickness and field strength respectively. The time scale is $L/U_A$, the Alfvén time scale.

The effects of turbulence enter into the model through the electomotive force $\emf=\langle\u'\times\b'\rangle$. We follow \cite{Higashimori13} in ignoring the effects of turbulent terms deriving from the the momentum equation,  which is in line with other theories of turbulent reconnection \citep[e.g.][]{Lazarian99}. While this choice can be motivated by the fact that we are studying reconnection in a magnetically-dominated plasma, we can show \emph{a posteriori} that the neglect of such terms is justified for the application in this paper. This analysis is presented in the Appendix. Also, this choice also allows us to compare more closely with previous work in this area \citep{Higashimori13,Widmer16b,Widmer16a,Widmer19}.

By means of the TSDIA approach, the electromotive force can be written in the form
\begin{equation}\label{emf}
	\emf = \alpha\B-\beta\J+\gamma\Om,
\end{equation}
where $\J=\nabla\times\B$ and $\Om=\nabla\times\U$. The transport coefficients $\alpha$, $\beta$ and $\gamma$ involve the one-point turbulent statistical quantities, which can be expressed as
\begin{align}
	\alpha &= C_\alpha \tau H,  &H=\left<\b'\cdot \cur' - \u'\cdot \mathbf{\boldsymbol\omega}'\right>,\\
	\beta &= C_\beta \tau K,  &K=\left<\u'\cdot \u' + \b'\cdot \b'\right>,\label{beta}\\
	\gamma &= C_\gamma \tau W,  &W=\left<\u'\cdot \b'\right>.
\end{align}
The TSDIA determines these transport coefficients by assuming the turbulent statistics come from fully-developed small-scale turbulence that decays over a turbulent time scale $\tau$.  This time scale is a constant that is chosen depending on the application. The model constants $C_{\alpha}$, $C_{\beta}$ and $C_\gamma$ are universal constants derived from the TSDIA approach and have the values $C_\alpha= 0.01$ and $C_\beta$, $C_\gamma = 0.3$ \citep{Higashimori13,Yokoi2020}.

The first transport coefficient, $\alpha$, is related to the turbulent helicity of the system. This term has received significant attention in dynamo studies in relation to the so-called $\alpha$-effect. Due to the symmetry of the reconnection problem that we will focus on, the $\alpha$ term is negligible and, like in \cite{Higashimori13}, we ignore it.

The remaining two transport coefficients, as shown by \cite{Higashimori13}, do play an important role in reconnection. The $\beta$ term is related to the turbulent energy $K$ and can be responsible for either enhancing or dissipating the mean magnetic field. The $\gamma$ term is related to the turbulent cross-helicity $W$. Again, this term can lead to the suppression or enhancement of the magnetic field but primarily determines the spatial structure of the turbulent energy.  In particular for 2D null point reconnection, cross-helicity has a quadrupolar structure which helps to concentrate the turbulent energy at the null point. \cite{Higashimori13} showed that without cross-helicity, a slightly less-concentrated diffusion region forms, resulting in less flux being reconnected compared to when cross-helicity is included in the model. In the diffusion region, however, it is the turbulent energy $K$ that dominates.

To complete the TRM, evolution equations are required for the one-point turbulent statistical quantities $K$ and $W$. These are derived to be
\begin{align}
	&\frac{\partial K}{\partial t} + \U \cdot \nabla K = -\emf\cdot\J + \B\cdot\nabla W - \epsilon_K, \label{K_eqn} \\
	&\frac{\partial W}{\partial t} + \U \cdot \nabla W = -\emf\cdot\Om + \B\cdot\nabla K - \epsilon_W, \label{W_eqn}
\end{align}
where $\epsilon_K$ and $\epsilon_W$ are the dissipation terms of $K$ and $W$ respectively. A model for the turbulent dissipation is the final requirement. We adopt the same approach as \cite{Higashimori13} and model these quantities as
\begin{equation}
	\epsilon_K = \frac{K}{\tau}, \quad \epsilon_W = C_W\frac{W}{\tau},
\end{equation}
where $C_W=1.3$  and $\tau$ is the time scale of turbulence. We adopt this approach in order to investigate all possible solutions of the TRM and not only the solutions of \cite{Widmer19}.

This presentation of the TRM equations mirrors that from \cite{Higashimori13} in all aspects but one.  The mean field equations in \cite{Higashimori13} are compressible but they assume that the turbulence is incompressible (the electromotive force \ref{emf} results from the TSDIA approach while assuming incompressibility). We focus entirely on incompressible MHD at all scales.

\section{Flux and field line conservation in the TRM}
When discussing reconnection and topology in the context of the TRM, some care is required. Unlike in laminar MHD, only the \emph{mean} field quantities are described explicitly. Thus, when we speak of reconnection in the TRM, it is reconnection of the mean magnetic field and not the full (mean plus fluctuating) field that is described. It is in this sense that we can speak of \emph{ideal reconnection}, as turbulence can cause the reconnection of mean field lines without the explicit need of imposed magnetic diffusion. We now explore this in more detail by considering flux and field line conservation (for the mean magnetic field). 

\subsection{Flux conservation}
Alfvén's celebrated result for ideal MHD \citep{Alfven42} shows that magnetic flux is frozen into the flow and, as a consequence of this, field line topology is preserved (field lines behave as material lines). For the TRM, there are similarities to this situation, but also fundamental differences.

Let us assume that the mean velocity and magnetic fields are suitably smooth (possessing as many derivatives as needed - a reasonable assumption for the mean fields). Let $\curlyC$ be a closed loop, with unit tangent vector $\t$, that is (Lie-)transported by the mean velocity $\U$. Let the corresponding surface, bounded by $\curlyC$, be denoted $\curlyS$ and let the unit normal vector to this surface be denoted $\n$. The mean magnetic flux $\Phi$ through $\curlyS$ is
\[
\Phi = \int_\curlyS\B\cdot\n\,\d^2x.
\]
It is not difficult to show that, with magnetic diffusion ignored,
\begin{align*}
    \frac{\d\Phi}{\d t} &= \int_\curlyS\left[\frac{\partial\B}{\partial t} - \nabla\times(\U\times\B)\right]\cdot\n\,\d^2x\\
    &=\oint_\curlyC\emf\cdot\t\,\d x,
\end{align*}
where the last line follows from an application of Stokes' theorem. It is, therefore, clear that the mean magnetic flux is conserved if the electromotive force has the form $\emf=\nabla\phi$, for some scalar function $\phi$. In this situation, we have a modified form of Alfvén's theorem, though there are important physical differences. For simplicity, let us consider the case of $\phi={\rm const.}$, i.e. $\emf=\boldsymbol{0}$. Superficially, this choice seems to satisfy flux conservation trivially, akin to removing magnetic diffusion from the induction equation. However,  on consideration of equation (\ref{emf}) for this case, namely
\begin{equation}\label{emf0}
\boldsymbol{0}=\alpha\B-\beta\J+\gamma\Om,
\end{equation}
this situation is more complex. First, the trivial solution $\alpha=\beta=\gamma=0$ means that there is no turbulence and we are back to laminar MHD. If the turbulent transport coefficients are non-zero, however, equation (\ref{emf0}) represents a \emph{dynamic balance} \citep{Yokoi2020}. This means that although the mean flux is conserved, it is not simply in a passive manner as for the magnetic field in laminar ideal MHD, in which the full, rather than just the mean, magnetic flux is frozen into the flow. Instead, turbulence acts continuously in a particular way (satisfying equation \ref{emf0}) to conserve the mean flux. Thus, although the derivations of flux conservation in laminar ideal MHD and mean flux conservation in the TRM are very similar, the underlying physics of both is markedly different.   

 In general, the electromotive force is not likely to simply be zero or a potential field (we will show such an example later), so the effects of turbulence will violate Alfvén's theorem for mean fields. Thus, the mean magnetic flux is not, in general, conserved in the TRM.

\subsection{Field line topology}
For laminar ideal MHD, a corollary of Alfvén's theorem is that the topology of field lines is also conserved. Translating this result to the TRM, the condition for the conservation of (mean) field line topology can be written as
\[
\frac{\partial\B}{\partial t} - \nabla\times(\U\times\B) = \lambda\B,
\]
for some scalar function $\lambda$ \citep{1966SSRv....6..147S,Hornig96}. We thus have more options for mean field line conservation than just setting $\emf=\nabla\phi$, which corresponds to setting $\lambda=0$. In other words, the mean field line topology can be conserved even if the mean flux is not. We now follow a standard approach (see, for example, section 2.2 in \cite{Birn07} by Hornig) but applied to the mean fields. Let us assume that there exists a velocity field $\V$ such that
\[
\frac{\partial\B}{\partial t} = \nabla\times(\V\times\B).
\]
That is, the mean magnetic flux is frozen into the flow determined by $\V$. The ideal and mean induction equation for a general electromotive force is
\[
\frac{\partial\B}{\partial t} = \nabla\times(\U\times\B)-\nabla\times\emf,
\]
where we assume no dynamic balance. Equating these two induction equations, it is clear that
\[
\nabla\times[(\U-\V)\times\B] = \nabla\times\emf.
\]
If we write $\W=\U-\V$, then for the electromotive force to preserve the mean field line topology, it must have the form
\begin{equation}\label{emf_top_preserve}
\emf = \W\times\B+\nabla\phi.
\end{equation}
The individual parts of $\emf$ in equation (\ref{emf}) do not have the form of the first or second terms on the right-hand side of equation (\ref{emf_top_preserve}). Therefore, in general, the effect of turbulence is to not preserve mean field line topology but to cause reconnection. In particular, the mathematical form of the term related to the turbulent energy in equation (\ref{emf}) is very similar to that of the non-ideal term of resistive MHD (magnetic diffusion), even though the physics represented by the two terms is different. Thus, in general, turbulence leads to the reconnection of mean field lines. 

These general theoretical considerations imply that reconnection in the TRM depends on the dominance of one or more of the terms that compose the electromotive force. The efficiency of reconnection depends on how well $|\emf|$ of sufficient size can be localized, similar to the localization of magnetic diffusion at magnetic topological boundaries (e.g. separators and null points) for reconnection in laminar resistive MHD. We thus have a mathematical correspondence between laminar and turbulent reconnection, even if the equations represent different physics. This suggests that for the phenomena of laminar reconnection can find turbulent analogues within the context of the TRM.  We now investigate this claim for 2D steady-state reconnection in a current sheet.

\section{Current sheet reconnection}
\subsection{Setup}
We solve the TRM equations numerically using the Method of Lines solver \texttt{Bout++} \citep{Dudson09}, a semi-discrete simulation framework that solves systems of PDEs in general curvilinear coordinates.  In our simulations, we discretize the spatial coordinates $(z,x)$ into a $N_z\times N_x$ uniform mesh in a simulation domain of (non-dimensional) size $L_z \times L_x$ (the $z$-direction is horizontal and the $x$-direction is vertical).  We solve the momentum and induction equations in stream-function form, so the full system of equations is
\begin{align}
    \frac{\partial}{\partial t} \nabla^2 \phi &= \left\{\nabla^2 \phi, \phi  \right\} + \left\{ \psi, \nabla^2 \psi \right\} + \frac{1}{\Re}  \nabla^4 \phi, \\
    \frac{\partial}{\partial t} \psi &= \left\{\psi, \phi  \right\} + \left( C_\beta \tau K + \frac{1}{\Rm} \right)  \nabla^2 \psi - C_\gamma \tau W \nabla^2 \phi, \\
    \frac{\partial}{\partial t} K &= \left\{K, \phi  \right\} + \left\{\psi, W  \right\} +C_\beta \tau K \left[ \nabla^2 \psi \right]^2 - C_\gamma \tau W \nabla^2 \psi \nabla^2 \phi \nonumber \\
    & - \frac{K}{\tau} - \chi \nabla^4 K, \\
        \frac{\partial}{\partial t} W &= \left\{W, \phi  \right\} + \left\{\psi, K  \right\} +C_\beta \tau K \nabla^2 \psi \nabla^2 \phi - C_\gamma \tau W \left[ \nabla^2 \phi \right]^2 \nonumber \\
    & - C_W\frac{W}{\tau} - \chi \nabla^4 W, 
\end{align}
where $\U = \nabla \times  ( \phi {\boldsymbol e_y} )$ and $\B=\nabla \times  ( \psi {\boldsymbol e_y} )$. We have included hyper diffusivity on the turbulent quantities for stability with a hyper-diffusion coefficient of $\chi=10^{-7}$. The curly braces are Poisson brackets defined as 

\[\left\{ A,B \right\}= \frac{\partial A}{\partial x}\frac{\partial B}{\partial z}-\frac{\partial B}{\partial x}\frac{\partial A}{\partial z}.\]  
The brackets are discretized via Arawaka's Method \citep{Arakawa66} while the laplacian terms are solved by 2nd order central differences. This reduces the entire problem to a system of ODEs that are then integrated in time via the implicit \texttt{pvode} scheme, a backwards differencing method \citep{Byrne99}.
The initial form of the magnetic field in all simulations takes the form of a  Harris current sheet
\begin{align}
    \B_0 = \tanh\left(\pi \frac{x}{\delta}\right) \e_z,
\end{align}
where $\delta$ is the half-thickness of the current sheet. There is no initial flow and the initial turbulent energy is set to $K_0 = K_{min}$ , the smallest value the turbulent energy can take in the simulations. We set this minimum value to $K_{min}=10^{-8}$, which is as small as floating point precision allows.  The initial condition for $K_0=K_{min}$ then corresponds to a laminar current sheet. 

The turbulent time scale is determined as in \cite{Higashimori13}, in which we first consider the dominant balance of terms in the steady-state form of equation (\ref{K_eqn}). If $\J_0$ is the initial current density and $J_0$ represents its magnitude at the centre of the current sheet, for a steady-state we have
\begin{equation}\label{ss0_time_scale}
    \tau_0 \approx \frac{1}{\sqrt{C_\beta}J_0}.
\end{equation}
The turbulent time scale is varied from its steady-state value by rescaling as $\tau=C_\tau\tau_0$. We will make use of the scale factor $C_\tau$ later when characterizing reconnection solutions.
\par
In the following simulations, the current sheet (half-)thickness is set as $\delta=1$.  Unless otherwise specified, the lengths of the domain are, $L_x=10$ and $L_z=80$.  Our simulations are run at a resolution of $1024\times 2048$ with  periodic boundary conditions in $z$-direction and perfectly conducting, no slip conditions in the $x$-direction. The background Reynolds numbers  (fluid and magnetic) are both set to $5 \times 10^{4}$.
\par
To perturb the current sheet, we introduce a mix of modes that leads to reconnection at the centre of the domain. In terms of the magnetic flux function, the perturbation has the form
\begin{equation*}
\phi_0= \frac{1}{1000}\: \exp[-100x^2] \sum_{n=1}^{13} \frac{1}{n}\cos(nz).
\end{equation*}
This expression is designed to produce a small-amplitude perturbation that is limited to near the centre of the current sheet.

\subsection{Reconnection rates}

In order to determine whether or not a particular reconnection solution is fast, we need to determine the rate of reconnection.  There are various ways to measure the reconnection rate but the standard method, particularly for 2D reconnection, is to measure the Alfvén Mach number at the edge of the current sheet, $M_{\rm in}:= {U_{x,{\rm in}}}/{U_{A,{\rm in}}}$.  Theories of steady-state reconnection provide expressions for the reconnection rate based on particular scalings of the diffusion region. For example, in the Sweet-Parker scaling for resistive MHD, the reconnection rate is given by
\begin{equation}
    M_{\rm in} \sim  \frac{1}{\sqrt{ \Rm}} = \sqrt{\eta},
\end{equation}
where we introduce the $\eta$-notation for the constant background magnetic diffusion coefficient, in order to simplify notation later.
\begin{figure}
    \centering{\includegraphics[width=0.8\textwidth]{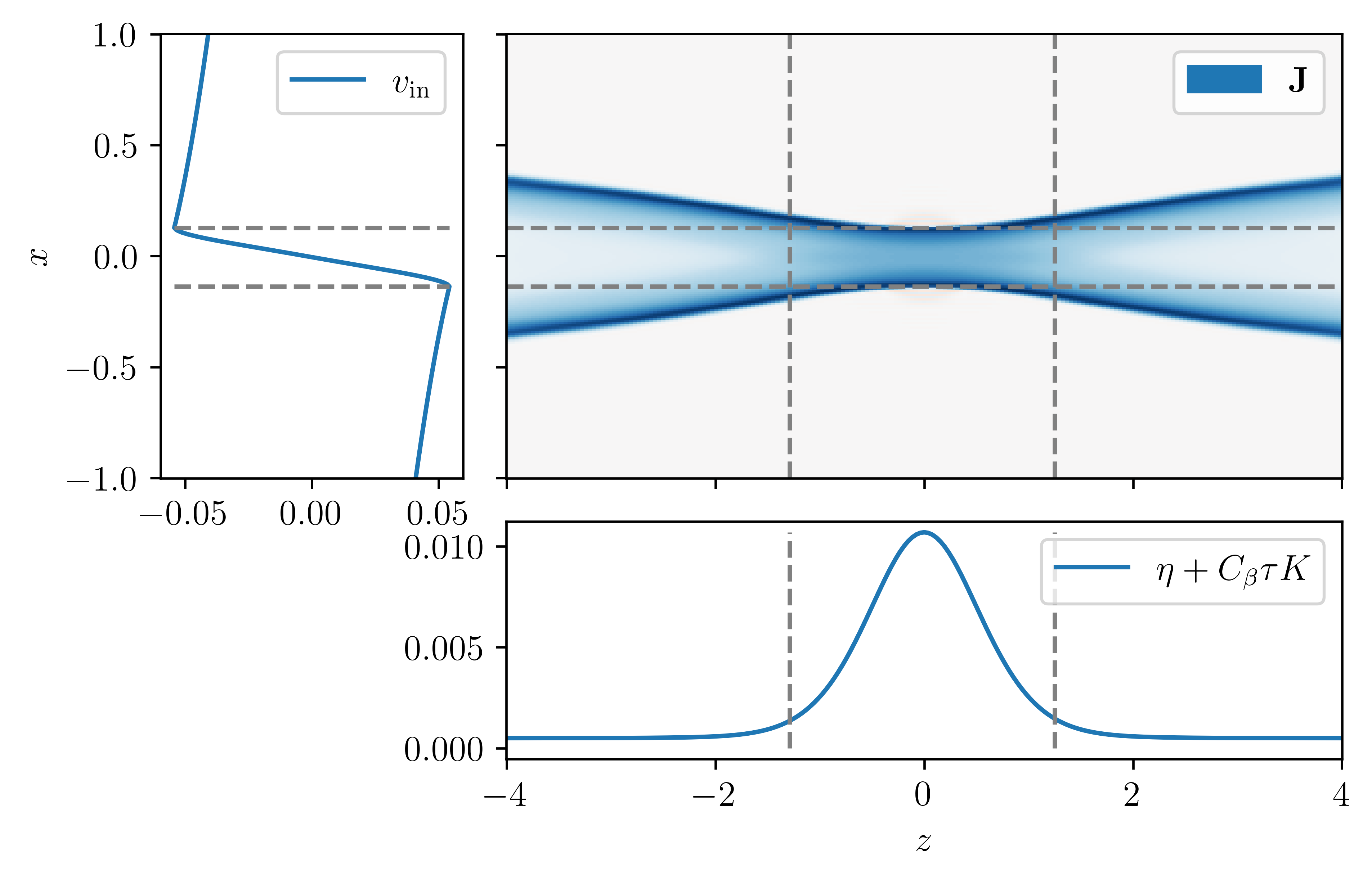}}
    \caption{A representation of the identification of the diffusion region.  The central figure shows a picture of the current density magnitude $J$ for a simulation with $C_\tau=1$.  The left panel shows in inflow velocity across the center of the sheet and the bottom panel displays the magnitude of $\eta_{\rm eff}$.  The dotted lines correspond to where we identify the boundaries of the diffusion region to be, as described in the main text. }\label{fig:diffusion_region_boundaries}
\end{figure}

In our simulations, the size and shape of the diffusion region depends on the dominant physics controlling it, and we will describe this in more detail later. However, for the practical identification of the diffusion region's boundaries, we identify characteristics that apply to all the turbulent solutions under study. The height of the diffusion region is determined by the rapid change in the mean velocity, as shown in Figure~\ref{fig:diffusion_region_boundaries}. For the width, we select the range in which the \emph{effective magnetic diffusivity} $\eta_{\rm eff}$ is more than twice the average over the entire length of the current sheet. In order to define the effective magnetic diffusivity, we combine the effects of magnetic diffusion and turbulent energy from the electromotive force to identify
\begin{equation}\label{eta_eff}
\eta_{\rm eff} = \eta + C_\beta\tau K.
\end{equation}
This quantity can be derived by expanding the induction equation (\ref{induction}) using equations (\ref{emf}) and (\ref{beta}) and ignoring the contribution of cross-helicity, which is small in the diffusion region. It is clear from Figure~\ref{fig:diffusion_region_boundaries} that this approach selects a reasonable choice for the width of the diffusion region.

Since we solve an initial value problem, we never produce a perfect steady-state (and so, in all later discussion, the term \emph{steady-state} refers to \emph{quasi-steady-state}, for which there is weak decay with time). In practice, this means that the diffusion region changes in size by a small amount as the total magnetic flux is depleted over time. Thus, all of our reconnection measurements are based on an average of the five closest grid cells to the diffusion region boundary. 

\begin{figure}
    \centering{\includegraphics[width=0.7\textwidth]{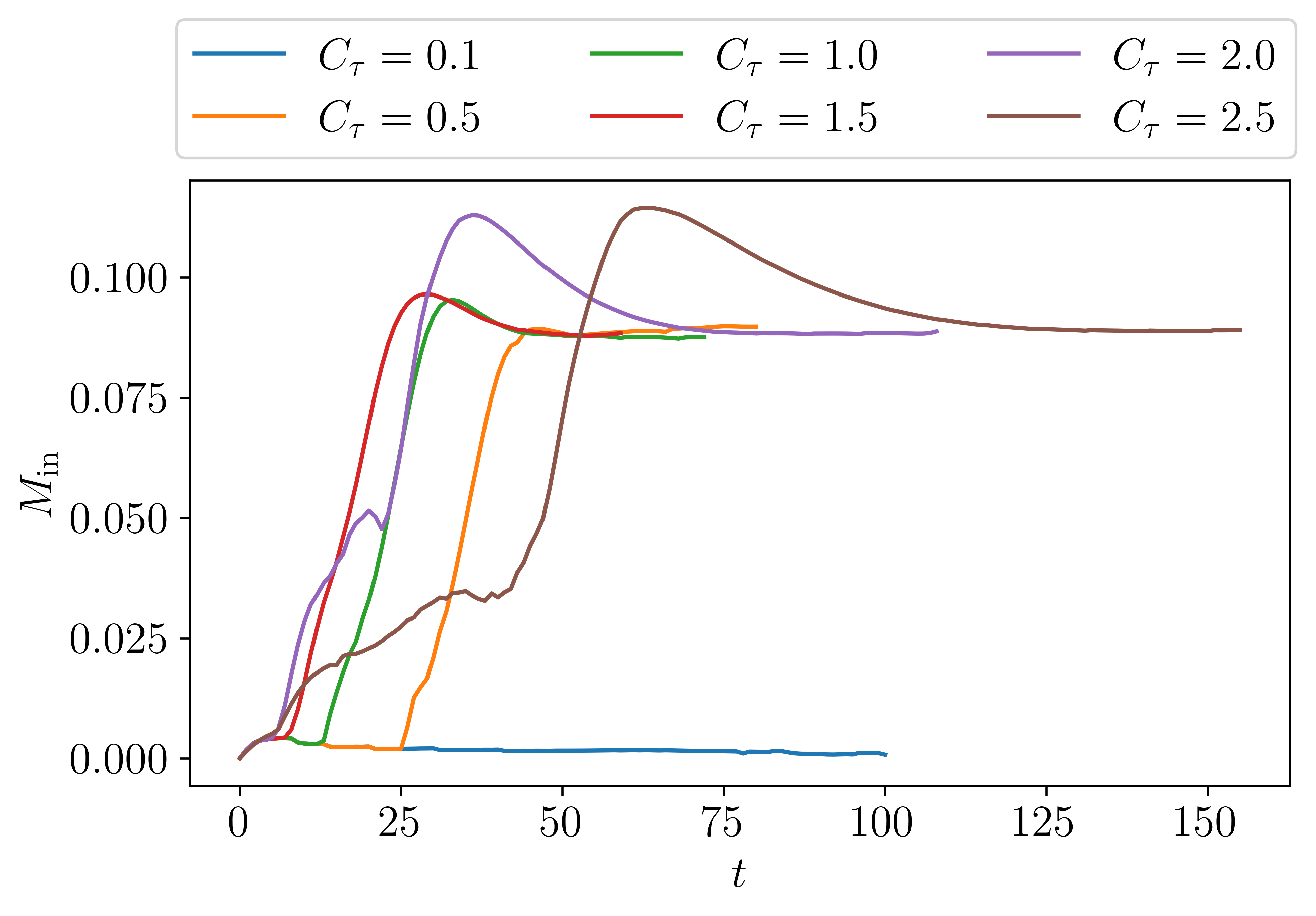}}
    \caption{This figure shows the reconnection rate $M_{\rm in}$ for various values of $C_\tau$.}\label{fig:reconnection_rate}
\end{figure}

Figure~\ref{fig:reconnection_rate} shows the reconnection rate of the TRM over time for various values of the turbulent time scale factor $C_\tau$ (equivalent to considering different $\tau$). What is clear is that for values of $C_\tau$ large enough, the reconnection rate evolves to a unique steady rate just below the fast rate of $M_{\rm in}\approx 0.1$ \citep{Cassak17}. For lower values of $C_\tau$ the steady-state reconnection rate remains low, typically less than 0.01. Remarkably, for this complex system, there are only two general types of steady reconnection solution. To justify the existence of these two steady solutions, we will identify the critical balances that lead to their formation. Before this, however, we now present an overview of the nature of all reconnection solutions.

\subsection{Overview of all solutions}
\begin{figure}
    \centering{
    \begin{subfigure}{0.45\textwidth}
        \includegraphics[width=\textwidth]{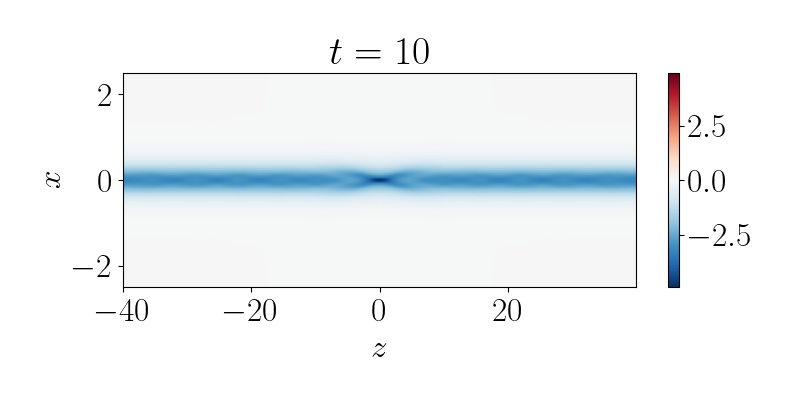}
        \caption{$C_\tau=0.1$}
    \end{subfigure}
    \begin{subfigure}{0.45\textwidth}
        \includegraphics[width=\textwidth]{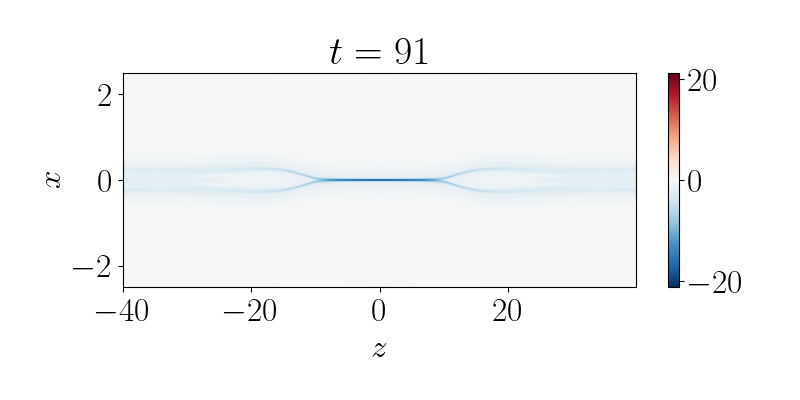}
        \caption{$C_\tau=0.1$}
    \end{subfigure}\\
    \begin{subfigure}{0.45\textwidth}
        \includegraphics[width=\textwidth]{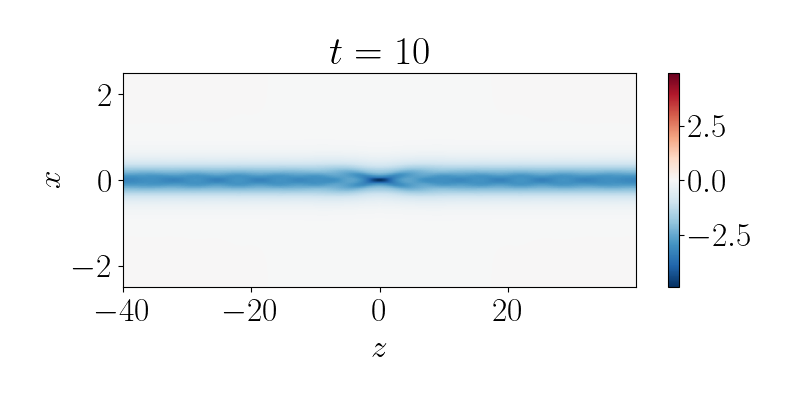}
        \caption{$C_\tau=1.0$}
    \end{subfigure}
    \begin{subfigure}{0.45\textwidth}
        \includegraphics[width=\textwidth]{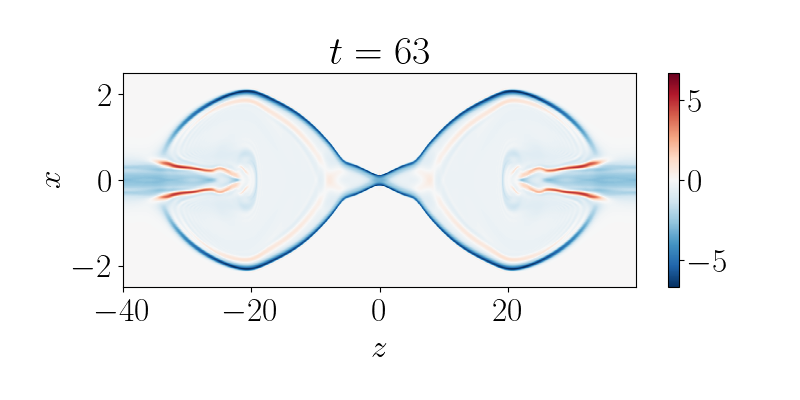}
        \caption{$C_\tau=1.0$}
    \end{subfigure}\\
    \begin{subfigure}{0.45\textwidth}
        \includegraphics[width=\textwidth]{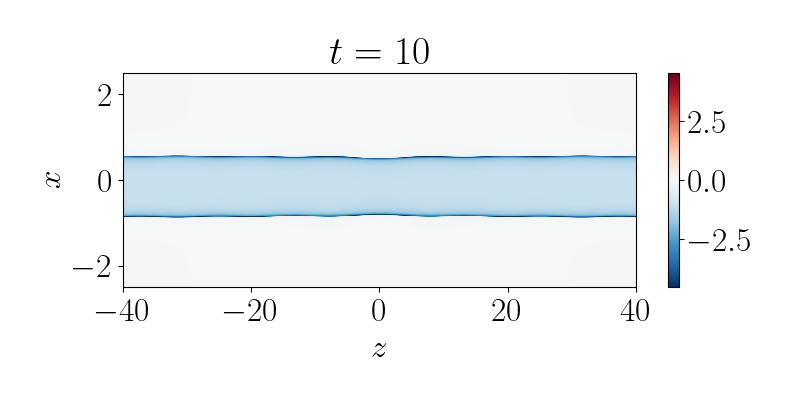}
        \caption{$C_\tau=2.5$}
    \end{subfigure}
    \begin{subfigure}{0.45\textwidth}
        \includegraphics[width=\textwidth]{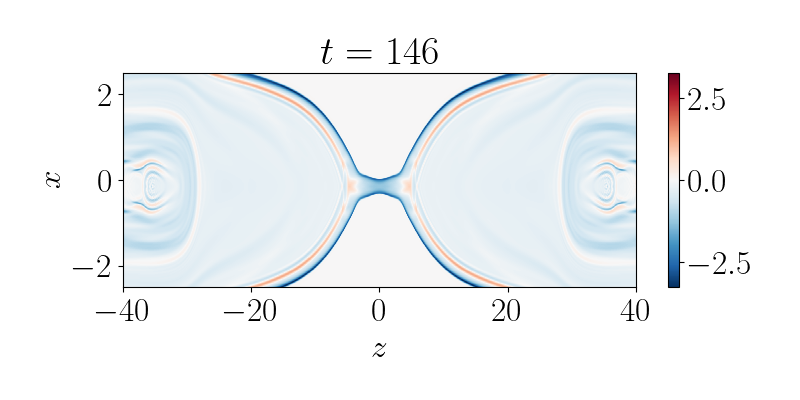}
        \caption{$C_\tau=2.5$}
    \end{subfigure}\\
    }
    \caption{Maps of the current density magnitude $J$ displaying different phases of the reconnection solutions for different values of $C_\tau$.  The images on the left (a, c, e) depict an early stage of reconnection, before significant deformation of the current sheet. The images on the right 
    (b, d, f) depict when steady-state reconnection has been established.  Note that we have zoomed in on the diffusion region so $x \in [-2.5,2.5]$.  The domain for the $C_\tau=2.5$ case has been extended to $L_z=160$ while retaining the same resolution so that a steady-state can be achieved before the outflow impinges upon the boundary.}\label{fig:current_sheet_pictures}
\end{figure}

As mentioned earlier, it has been reported that three distinct reconnection solutions exist for the TRM model: laminar reconnection, turbulent reconnection, and turbulent diffusion. Here we show that only the first two of these solutions are distinct, with the third solution (turbulent diffusion) revealing itself as a version of the second (turbulent reconnection) on a long time scale. Representations of these solutions are displayed in Figure \ref{fig:current_sheet_pictures}.

Focussing on each solution in turn, Figures \ref{fig:current_sheet_pictures} (a) and (b) display initial and late times of laminar reconnection. This solution is laminar in the context of the TRM as there is no growth of turbulent energy $K$ (to be discussed in more detail shortly) and the current sheet evolves to the classic Sweet-Parker scenario of resistive MHD.

For the other solutions, shown in Figures \ref{fig:current_sheet_pictures} (c) and (d) and (e) and (f) respectively, the form of reconnection is governed by turbulence. Comparing the early phases in (c) and (e), the solution with large $C_\tau$ has led to a much thicker and diffuse current sheet. The initial reconnection rate for this $C_\tau=2.5$ case is much smaller than that of the $C_\tau=1$ case on the time scale which the latter develops fast reconnection (see Figure \ref{fig:reconnection_rate}). This slower reconnection has been previously described as turbulent diffusion \citep{Higashimori13}. However, if the solution is allowed to evolve to much later times, it develops the same fast reconnection as for smaller $C_\tau$ (turbulent reconnection) cases.

The magnetic field of the fast turbulent reconnection solutions follows the structure of Petschek reconnection, as opposed to laminar Sweet-Parker reconnection. As is clear from  Figures \ref{fig:current_sheet_pictures} (d) and (f), the diffusion region is localized in the centre of the domain with four thin current layers\footnote{Although slow-mode shocks feature in the standard description of Petschek reconnection, we describe these as \emph{current layers} since we are solving the \emph{incompressible} MHD equations.} emanating diagonally outwards. Beyond this visual inspection, we also present evidence that the localized diffusion region behaves approximately like a Sweet-Parker region with a local reconnection rate 
\[
M_{\rm in}\sim \Rm_{\rm eff}^{-1/2}, 
\]
where the localized magnetic Reynolds number at the diffusion region is $\Rm_{\rm diff} = L U_{A,{\rm in}}/\eta_{\rm eff}$ \citep[e.g.][]{Baty06}. Table \ref{tab:scalings} shows a close match for the Sweet-Parker scaling across a range of $C_\tau$.  

\begin{table}
\begin{center}
\caption{A comparison of the Mach number at the edge of the diffusion region, $M_{\rm in}$, and the Sweet-Parker reconnection rate in the diffusion region, $\Rm_{\rm eff}^{-1/2}$, for a range of values of $C_\tau$. The values are taken from a time average starting at $t_\text{sim}=t_\text{end}-40$ and proceeding to the end of the simulation, $t_\text{sim}=t_\text{end}$.  The error bars come from the error in measuring the inflow region across five grid cells of the boundary, as mentioned in the main text.}\label{tab:scalings}
{\begin{tabular}{l | cccc}
$C_\tau$& $0.5$  & $1.0$  & $1.5$  & $2.0$ \\
\hline
\\
$M_{\rm in}$& $0.092 \pm 0.003$  & $0.089 \pm 0.002$  & $0.090 \pm 0.002$  & $0.090 \pm 0.001$ \\
$\Rm_{\rm eff}^{-1/2}$& $0.079 \pm 0.001$  & $0.089 \pm 0.001$  & $0.096 \pm 0.001$  & $0.094 \pm 0.001$ \\[6pt]

\end{tabular}}
\end{center}
\end{table}
In summary, the TRM supports two steady-state reconnection solutions, a slow and laminar Sweet-Parker solution and a fast and turbulent Petschek solution. We now consider how these solutions are selected.

\subsection{Critical current balances}
The two steady-state reconnection solutions are due to intrinsic balances within the system. In searching for these balances, there are two key physical elements to consider. First, the inclusion of a constant background magnetic diffusivity $\eta$ leads to a natural length scale. In a steady-state current sheet, the classical Sweet-Parker scaling provides an estimate of the current sheet thickness as $\delta\sim \eta^{1/2}$. This estimate provides a lower length scale that can be reached by the system in a steady-state when the presence of turbulence is absent. For our purposes, it will be useful to provide an estimate of the current density magnitude at the reconnecting X-point. Under the Sweet-Parker scaling this quantity can be approximated as
\begin{equation}\label{jc_sp}
    J_\eta \sim \frac{B_{in}}{\eta^{1/2}},
\end{equation}
where $B_{\rm in}$ is the magnitude of the magnetic field at either of the upper or lower edges of the diffusion region.

\begin{figure}
    \centering{
    \begin{subfigure}{0.45\textwidth}
        \includegraphics[width=\textwidth]{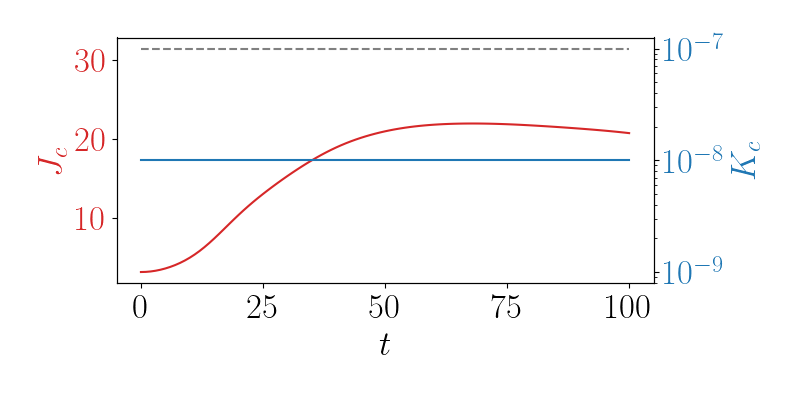}
        \caption{$C_\tau=0.1$}
    \end{subfigure}
    \begin{subfigure}{0.45\textwidth}
        \includegraphics[width=\textwidth]{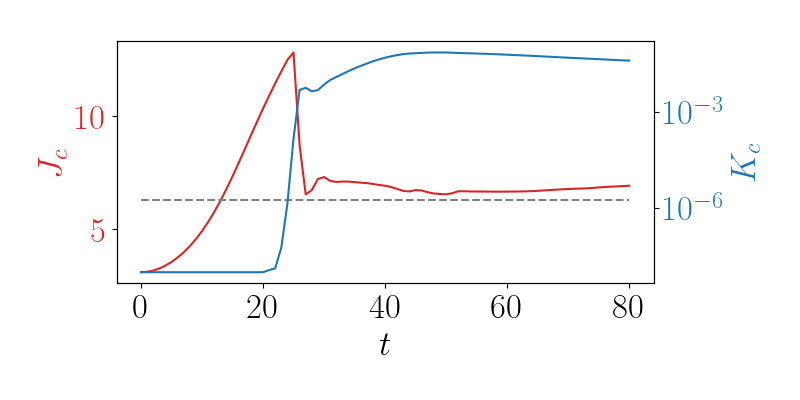}
        \caption{$C_\tau=0.5$}
    \end{subfigure}\\
    \begin{subfigure}{0.45\textwidth}
        \includegraphics[width=\textwidth]{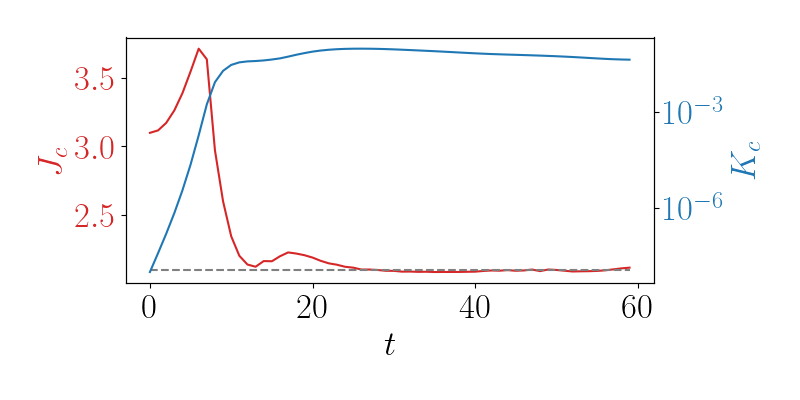}
        \caption{$C_\tau=1.5$}
    \end{subfigure}\\
    }
    \caption{Time evolutions of $J_c$ amd $K_c$ for different values of $C_\tau$. For each case, the value of $J_\tau$ is represented by a dashed line.}\label{fig:critical_balance}
\end{figure}

For a mean steady-state with turbulence, we look again to equation (\ref{K_eqn}). Similar to the derivation of equation (\ref{ss0_time_scale}), and now combining this with the expression $\tau=C_\tau\tau_0$, we arrive at the following estimate for the magnitude of the current density at the centre of the diffusion region,
\begin{equation}\label{jc_turb}
    J_c=J_\tau \sim \frac{J_0}{C_\tau},
\end{equation}
 where $J_\tau$ is the critical current density (this is made clear below). In order to examine the relevance of the estimates given in (\ref{jc_sp}) and (\ref{jc_turb}), we first consider how the magnitude of the current density at the centre of the diffusion region, denoted $J_c$, relates to these quantities. Figure \ref{fig:critical_balance} displays the temporal development of $J_c$ and $K_c$ (the value of $K$ at the centre of the diffusion region) for different values of $C_\tau$. The value of $J_0/C_\tau$ is displayed for each case as a horizontal dashed line.

Figure \ref{fig:critical_balance} (a) displays a case of Sweet-Parker reconnection ($C_\tau=0.1$). The current density $J_c$ increases until shortly after $t=50$ and then decreases at a slow rate (this is the quasi-steady Sweet-Parker solution). There is no growth in the turbulent energy $K_c$. One feature that is clear from this figure is that $J_c<J_\tau$ for the entire evolution. In Figure \ref{fig:critical_balance} (b), however, which displays a case of Petschek reconnection ($C_\tau=0.5$), $J_c$ breaks through the $J_\tau$ barrier. As the current density grows, it eventually (shortly after $t=20$ for this case) becomes strong enough to generate turbulent energy $K_c$ (see equation \ref{K_eqn}). The current density reaches a maximum value and then drops rapidly to a value just above $J_\tau$. Co-temporal with this rapid decrease, the turbulent energy $K_c$ rises rapidly until holding an approximately steady value. Beyond $t\approx30$, both $J_c$ and $K_c$ are approximately steady (this is the quasi-steady Petscheck solution). In Figure \ref{fig:critical_balance} (c), we again have a case of Petschek reconnection ($C_\tau=1.5$), but the difference to the previous case is that $J_c\gtrapprox J_\tau$ for the entire evolution. Here, the current density is strong enough to cause $K_c$ to increase from the initial condition. Later, the behaviour of $J_c$ and $K_c$ is qualitativly similar to the $C_\tau=0.5$ case, i.e. both achieve quasi-steady values with $J_c$ resting just above $J_\tau$. What the cases in Figure \ref{fig:critical_balance} show is that $J_\tau$ provides a critical barrier for the selection of a Petschek solution. If $J_c$ can become larger than $J_\tau$, then turbulent energy can be produced and a steady-state can form with $J_\tau$ acting as a lower boundary for the current density at the centre of the diffusion region.

\begin{figure}
    \centering{\includegraphics[width=0.8\textwidth]{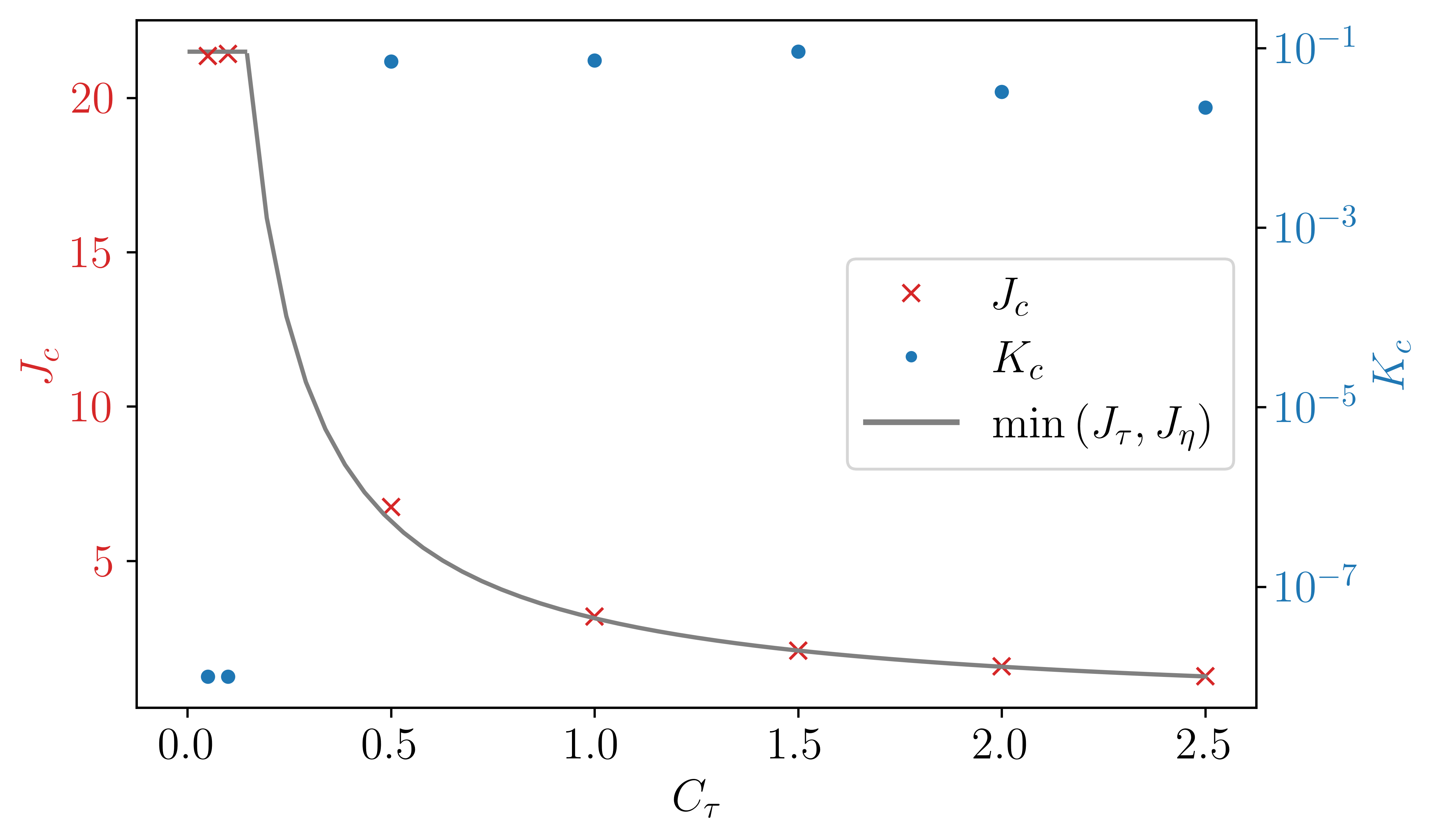}}
    \caption{A representation of the steady-state values of $J_c$ and $K_c$ for a range of $C_\tau$. The solid line shows the critical current selection $\min(J_\tau,J_\eta)$, which identifies the selection of either Sweet-Parker or Petschek solutions.}\label{fig:current_vs_tau}
\end{figure}

Figure \ref{fig:current_vs_tau} displays the relationship of the average the critical current balances $J_\eta$ and $J_\tau$ as a function of $C_\tau$, once a steady-state has been reached. There is clearly an excellent match between the simulation values of $J_c$, $J_\eta$ and $J_\tau$. For $C_\tau\gtrapprox 0.2$, $J_c\approx J_\tau$ and the turbulent energy reaches an approximately constant value of 0.1. These are all Petschek solutions. For $C_\tau\lessapprox 0.2$, $J_\tau$ is not reached and $J_c\approx J_\eta$. These are all Sweet-Parker solutions. 

\subsection{Self-similarity of turbulent Petschek solutions}

We have shown that for turbulent Petschek solutions, the steady reconnection rate is $M_{\rm in}\approx 0.1$ with $J_c\approx J_\eta$. Since $J_\tau\propto 1/\tau$, this suggests a \emph{self-similarity} for the turbulent Petschek solutions, all depending on the turbulent time scale. For example, compare the solutions displayed in Figures \ref{fig:current_sheet_pictures} (d) and (e). The latter has a value of $J_c$ 2.5 times less than that of the former. However, it is clear that the size of the diffusion region of the latter is greater than the former. Based on $J_\tau$, we can estimate the thickness of the diffusion region to be
\begin{equation}\label{delta}
    \delta\sim \frac{C_\tau B_{in}}{J_0}.
\end{equation}

\begin{figure}
    \centering{\includegraphics[width=0.8\textwidth]{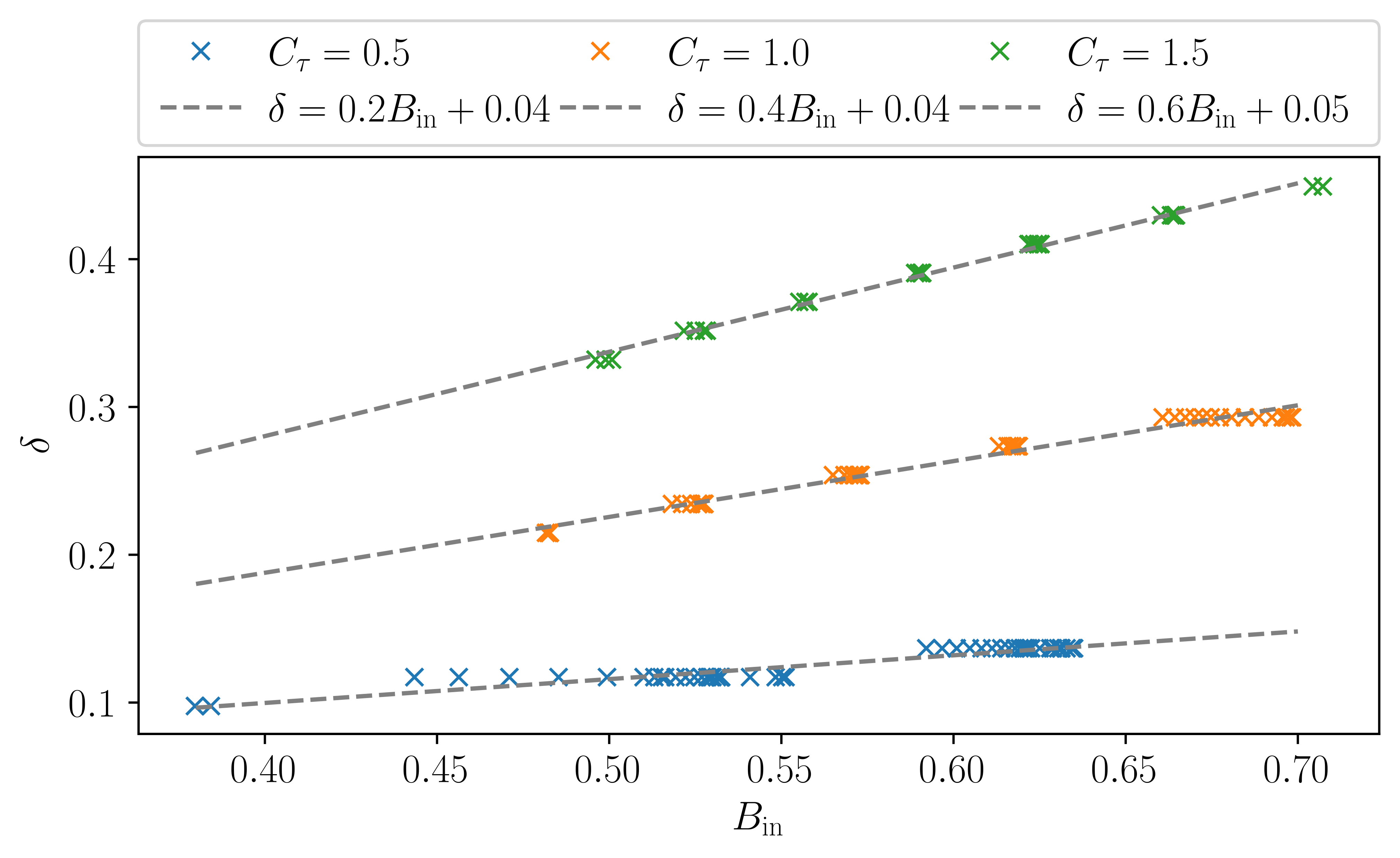}}
    \caption{Diffusion region thicknesses $\delta$, as a function of $B_{\rm in}$, for three values of $C_\tau$. Crosses are determined from simulations and lines of best fit are overplotted on the points of each case. The gradients of these lines follow the estimate in (\ref{delta}).   The discrete steps in the data are due to the resolution.  In the $x$-direction the resolution is $\Delta x \approx 0.01$ so, as the sheet decreases in width, $\delta$ decreases in integer multiples of $\Delta x$.}\label{fig:current_width}
\end{figure}

Figure \ref{fig:current_width} displays how the diffusion region thickness $\delta$ varies as a function of $B_{in}$ for three different values of $C_\tau$. For each case, a line of best fit is plotted over the points determined from the simulations. The gradient of each line is $C_\tau/J_0$, confirming the estimate in (\ref{delta}). Thus, the self-similarity of the turbulent Petschek solutions have diffusion region thicknesses and current densities scaling with $\tau$ and $1/\tau$ respectively. The solutions are self-similar, with the turbulent time scale acting as the fundamental scaling parameter. 

\subsection{Theoretical justification of the steady Petschek reconnection rate}
 Through the simulations described above, we have shown that a universal steady Petschek reconnection rate is possible and that different Petschek solutions are self-similar. We now provide some basic theoretical justification that a universal Petschek rate is a natural consequence of some standard balances combined with the behaviour of the turbulent energy at the null point.

First, let the localized diffusion region have a thickness and width denoted by $\delta$ and $L$ respectively. Let $U_{\rm in}$ denote the inflow speed, $U_{\rm out}$ denote the outflow speed and $K_c$ denote the turbulent energy at the null point, as before. From mass conservation, it follows that
\begin{equation}\label{mass_steady}
    U_{\rm in}L \sim U_{\rm out}\delta.
\end{equation}
Looking now to the momentum equation, balancing inertia with tension produces
\begin{equation}\label{mom_steady}
    U_{\rm out} \sim U_A.
\end{equation}
In the induction equation, it is the turbulent diffusivity $\beta_c\sim C_\beta\tau K_c$, which is important for reconnection. In the steady-state, we have the balance
\begin{equation}
    U_{\rm in} \sim \frac{\beta_c}{\delta} \sim \frac{C_\beta\tau K_c}{\delta}.
\end{equation}
Two more relations are required to determine the five independent variables, and these can be found by considering the behaviour of turbulence in the diffusion region. Close to the null point, the cross-helicity effects can be ignored. Further, from equations (\ref{ss0_time_scale}) and (\ref{jc_turb}), the critical current density can be written as $J_\tau = 1/(\sqrt{C_\beta}\tau)$. Combining these properties, it is a straightforward consequence that
\begin{equation}\label{diff_K}
    \frac{\partial K}{\partial t} +(\U\cdot\nabla)K \sim C_\beta\tau K(J^2-J_\tau^2).
\end{equation}
At the null (and stagnation) point, we have that $J=J_\tau$, which we have confirmed numerically. From this fact, scaling (\ref{delta}) was derived and can also be written as
\begin{equation}\label{delta2}
    \delta \sim \frac{B_{\rm in}}{J_\tau} \sim \sqrt{C_\beta}\tau B_{\rm in}.
\end{equation}
Close to the null point, the right-hand side of (\ref{diff_K}) may be ignored, and the turbulent energy is simply advected by the flow. At larger distances, $J\ll J_\tau$, representing the decay of turbulent energy. Given that the turbulent energy plays a controlling role in the definition of the diffusion region ($\beta_c\propto K_c$), the distance that an Alfvénic flow (for fast reconnection) travels in a decay time can be described as
\begin{equation}\label{Delta}
    L \sim U_A\tau.
\end{equation}
Putting all of this information together, the dominant scalings for steady Petschek reconnection can be written as the following five relations,
\begin{equation}
    L\sim U_A\tau, \quad \delta\sim\sqrt{C_\beta}\tau B_{\rm in}, \quad U_{\rm in} \sim \sqrt{C_\beta} U_A, \quad
    U_{\rm out} \sim U_A, \quad K_c\sim U_A^2.
\end{equation}
Thus, it follows that the steady Petschek reconnection rate is 
\begin{equation}\label{rrate_steady}
    M_{\rm in} = \frac{U_{\rm in}}{U_A} \sim \sqrt{C_\beta}.
\end{equation}
Since $C_\beta$ is a universal constant, it follows that the steady rate in (\ref{rrate_steady}) is also universal, matching qualitatively the behaviour found numerically.

\begin{figure}
    \centering
    \includegraphics[width=0.7\linewidth]{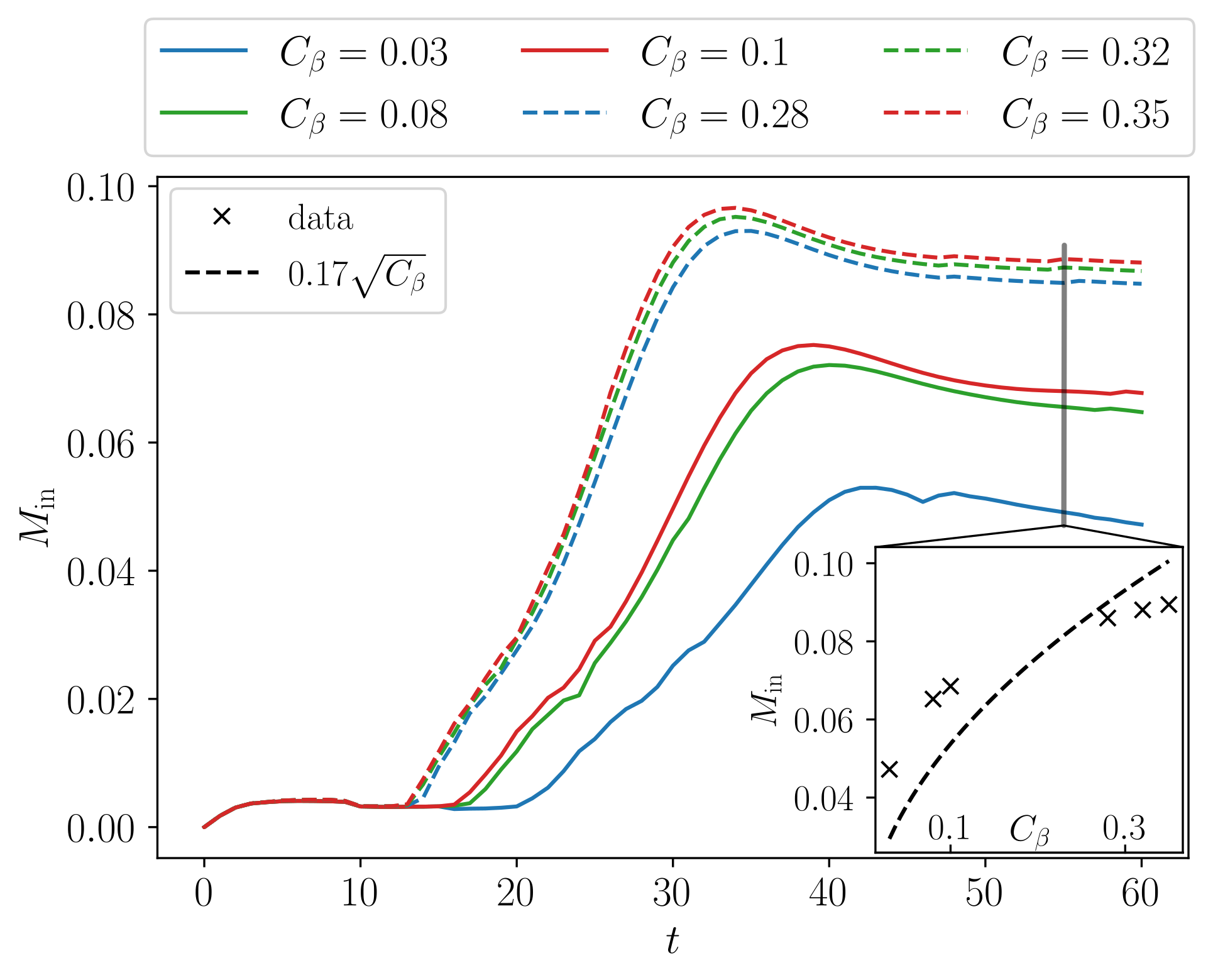}
    \caption{The behaviour of $M_{\rm in}$ as a function of the (given) parameter $C_\beta$. Note that $C_\gamma=C_\beta$ for each case. Here $C_\tau=1$}.
    \label{fig:cb_diff}
\end{figure}

Although $C_\beta$ is not a variable, we can confirm that, as suggested by (\ref{rrate_steady}), $M_{\rm in}$ follows approximately the above scaling. This feature is displayed in Figure \ref{fig:cb_diff} by considering ``high'' values of $C_\beta$, i.e. $O(0.3)$, and ``low'' values, i.e. $O(0.01-0.1)$.

Given the simplicity of the above scaling argument (i.e. purely local, ignoring the internal diffusion region structure, ignoring $W$, etc.), the agreement shown in Figure \ref{fig:cb_diff} is reasonable. Although the cross-helicity $W$ is missing from the scaling argument, it does play a role in Figure \ref{fig:cb_diff} as, for each case, $C_\gamma=C_\beta$. Since both of these parameters represent properties of MHD turbulence, they need to be varied together.

\section{Summary and discussion}
In this work, we have investigated the turbulent reconnection model (TRM) of \cite{Higashimori13} in order to provide a deeper insight into the nature of the reconnection solutions it supports. First, we have described what reconnection means within the context of the TRM. Secondly, we have shown that the TRM supports steady-state Sweet-Parker and Petschek reconnection. Sweet-Parker reconnection corresponds to laminar reconnection, when no turbulent energy develops, and behaves as in laminar resistive MHD. Petschek reconnection develops when a critical current density, based on the turbulent time scale, is reached, allowing for the growth of turbulent energy. The localization of this turbulent energy  concentrates the anomalous resistivity within the diffusion region. We show that there is a self-similarity between the turbulent Petschek solutions, based on the turbulent time scale.  The steady Petschek reconnection rate is shown to be universal, i.e. not dependent on any variables of the system like $\eta$, and this can also be argued from the consideration of basic balances within the main equations combined with the behaviour of the turbulent energy near the null point. In this model, fast and steady reconnection can occur without the explicit need for collisionless terms in Ohm's law. Indeed, the fastest reconnection generated, made possible by turbulence, is a phenomenon of ideal MHD.

Our results extend those of \cite{Higashimori13} in three ways. First, we show that steady-state reconnection solutions are possible, as described above. Secondly, we show that a slowly-reconnecting solution, previously identified as turbulent diffusion, is actually just the early stage of a solution of fast turbulent reconnection that takes longer to evolve. Thirdly, we identify the criterion for solution selection based on a critical current balance.

In relation to the affirmation of \cite{Widmer19} that, in their terminology, turbulent diffusion and laminar reconnection solutions are artifacts of choosing a constant turbulent time scale, we would suggest an alternative interpretation. We have already shown that turbulent diffusion is just an early stage of turbulent reconnection. However, for the laminar solution, we have shown that this develops when the magnitude of the  maximum current density in the current sheet fails to reach a critical value based on the turbulent time scale. Therefore, there is a clear physical explanation for laminar Sweet-Parker reconnection. However, the inclusion of a model equation for turbulent dissipation may only have access to a specific region of the full solution space and, thus, miss the Sweet-Parker solutions. Further, the inherent instability of Sweet-Parker current sheets \citep[e.g.][]{Loureiro07,pucci14,MacTaggart2020} may lead to a dynamical evolution in practice, rather than a steady-state. This topic would need to be investigated further to confirm if steady Sweet-Parker solutions can be categorically excluded.

 As mentioned in the Introduction, there are several methods of modelling turbulent reconnection. The approach adopted here is particularly useful for modelling turbulence in large systems for which it may not be possible to resolve turbulent fluctuations as in the direct numerical simulations of \cite{2009ApJ...700...63K} and \cite{2009MNRAS.399L.146L}. Despite the differences in the modelling approaches, however, the most important result is that each approach produces fast turbulence-driven MHD reconnection that in laminar MHD is only possible with the inclusion of an extended Ohm's law \citep[e.g.][]{Birn01} or by imposing some localized anomalous resistivity \citep[e.g.][]{Baty06}. 

The critical scalings that we have found allow for the size of the diffusion region region to be estimated \emph{a priori}. For example, when deciding on the resolution to be used in a simulation, the estimate (\ref{delta}) could be used to make sure that the diffusion region is adequately resolved. The sizes of the Petschek diffusion regions are typically thicker than the laminar Sweet-Parker sheets.

\section*{Acknowledgements}

 It is a pleasure to acknowledge the insightful comments of David Hosking, and to thank him for a seminal contribution to this work. We are also grateful for the very helpful and constructive conversations with Nobumitsu Yokoi. S.S. acknowledges support from a Maclaurin Scholarship from the University of Glasgow. D.M. acknowledges support from a Leverhulme Trust grant (RPG-2023-182), a  Science and Technologies Facilities Council (STFC) grant (ST/Y001672/1) and a Personal Fellowship from the Royal Society of Edinburgh (ID: 4282). 

\appendix
 
\section{The neglect of turbulence terms related to the Reynolds-Maxwell stress tensor}

In this work, we have followed \citep{Higashimori13} in neglecting the influence of turbulence due to terms from the momentum equation. Although this has been justified on the grounds that we are considering a magnetically-dominated plasma, we now provide evidence to justify the neglect of these terms in this study.

\begin{figure}
    \centering
    \includegraphics[width=0.45\linewidth]{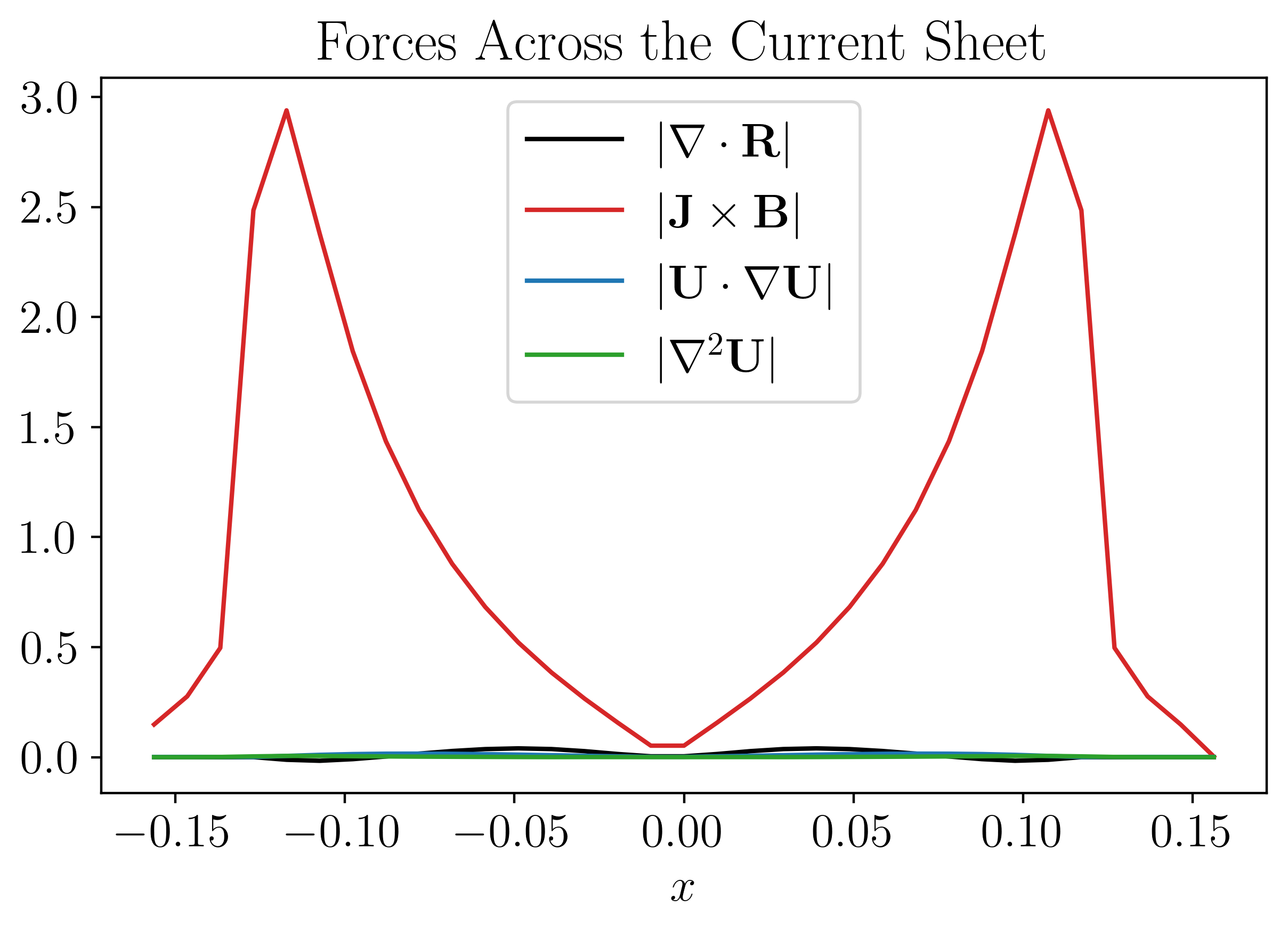}
    \includegraphics[width=0.45\linewidth]{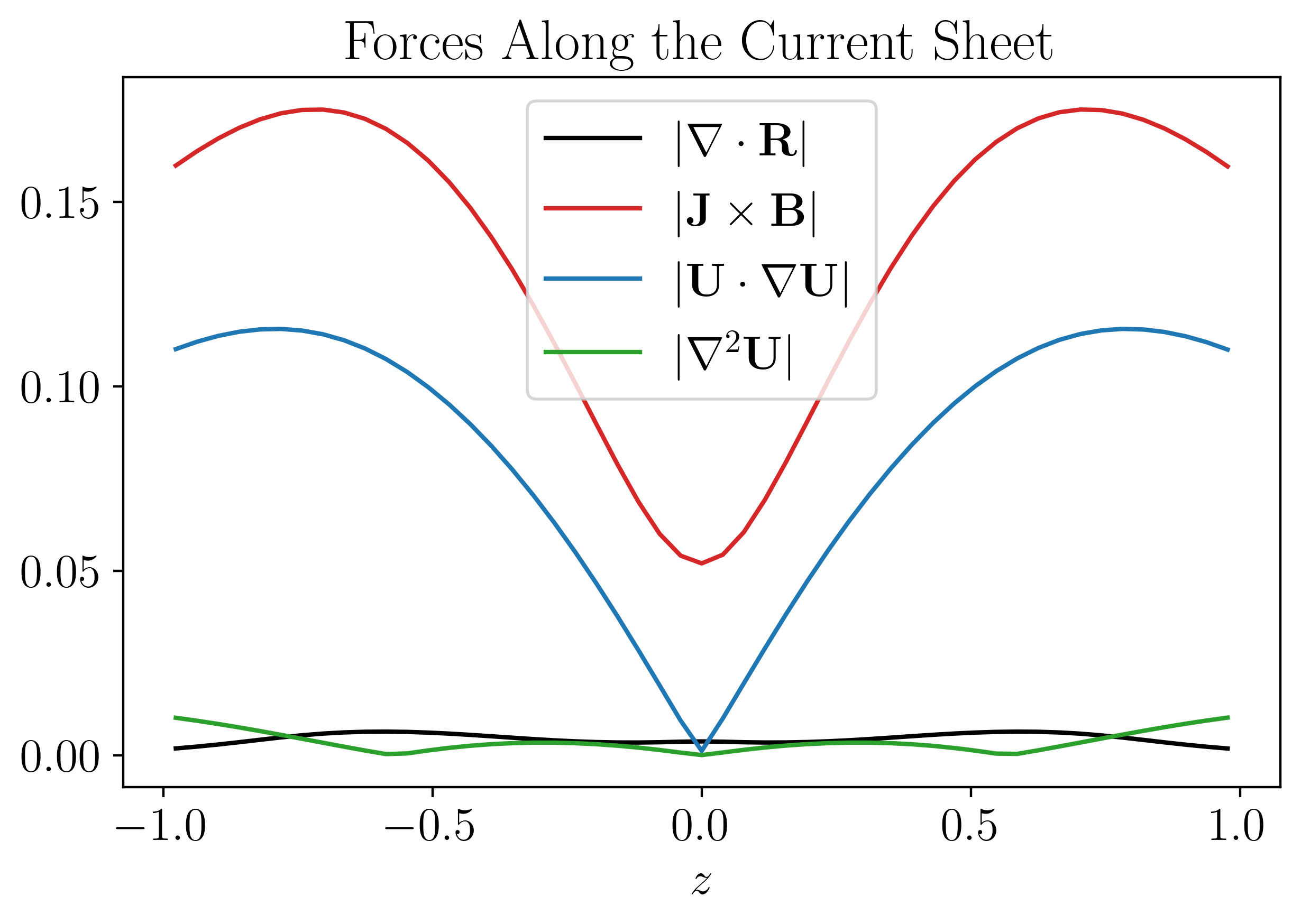}
    \caption{The magnitude of the Reynolds-Maxwell term, the Lorenz force, advection, and diffusion in the momentum equation both across (left) and along (right) the diffusion region.  In both cases, the effect of the Reynolds-Maxwell stress is small with respect to the other terms across the diffusion region. Here $C_\tau = 1$ and $t_{\rm sim}=63$}.
    \label{fig:reynolds_stress}
\end{figure}

First, if we were to write out the full mean momentum equation, we would have
\begin{equation}
    \frac{\partial\U}{\partial t} + (\U\cdot\nabla)\U = -\nabla\cdot\R -\nabla P + (\nabla\times\B)\times\B + \frac{1}{\Re}\nabla^2\U,
\end{equation}
where the new term, compared to equation (\ref{momentum}), is the divergence of $\R=\langle\u'\otimes\u'-\b'\otimes\b'\rangle$, the \emph{Reynolds-Maxwell stress tensor}. It is through this term that the effects of turbulence enter the momentum equation. After a considerable amount of algebra, the application of the TSDIA approach \citep[e.g][]{Yokoi2020} leads to the, leading order, representation of the Reynolds-Maxwell stress term as
\[
-\nabla\cdot\R = \nabla\cdot\left(\nu_K\S-\nu_M\M\right),
\]
where $\nu_K=\frac75\beta$, $\nu_M=\frac75\gamma$ and the remaining tensors are defined as
\[
\S = \nabla\U + (\nabla\U)^{\rm T}, \quad \M =  \nabla\B + (\nabla\B)^{\rm T}.
\]
For the reconnection problem considered in this work, Figure \ref{fig:reynolds_stress} displays the typical magnitudes of the forces at work.

It is clear from Figure \ref{fig:reynolds_stress} that the Reynolds-Maxwell stress does not play a dominant role in the force balances during the steady reconnection phase. 

\begin{figure}
    \centering
    \includegraphics[width=0.45\linewidth]{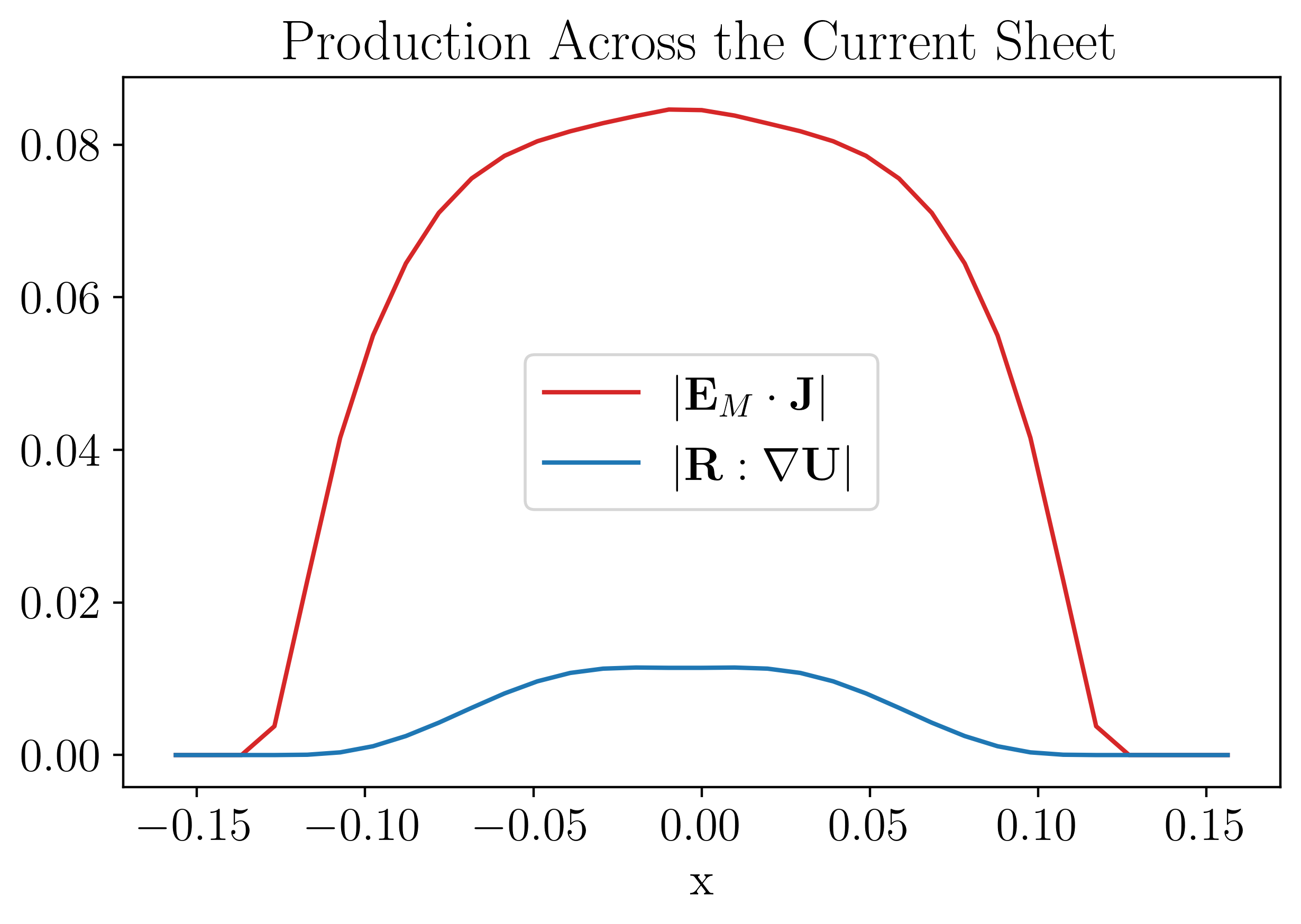}
    \includegraphics[width=0.45\linewidth]{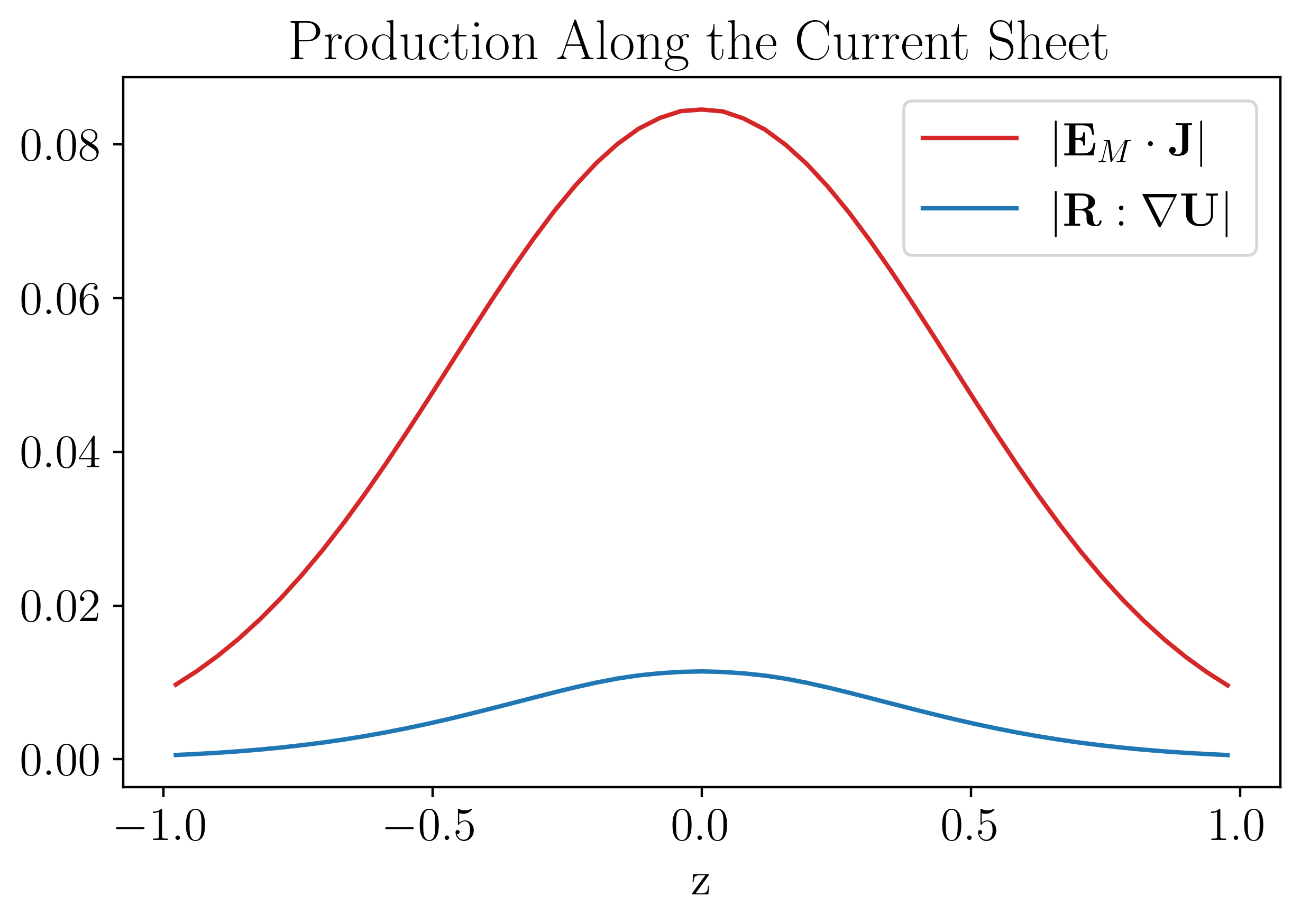}
    \caption{The magnitudes of the Reynolds-Maxwell and turbulent electromotive force production terms in the turbulent energy equation both across (left) and along (right) the diffusion region.  In both cases, the size of the production term due to the Reynolds-Maxwell stress is much less than that due to the electromotive force.  As above, $C_\tau = 1$ and $t_{\rm sim}=63$}.
    \label{fig:turbulent_production}
\end{figure}

More directly related to the turbulent reconnection is the production of the turbulent energy $K$. If we include the effects from the Reynolds-Maxwell stress tensor, equation (\ref{K_eqn}) becomes
\begin{equation}
    \frac{\partial K}{\partial t} + \U \cdot \nabla K = -\R:\nabla\U -\emf\cdot\J + \B\cdot\nabla W - \epsilon_K.
\end{equation}
The new term in the above equation does not play a significant role in the production of turbulent energy as indicated in Figure \ref{fig:turbulent_production}.

The results displayed here are typical throughout the period of steady reconnection. We stress, however, that this approximation is suitable for the configuration that we study here. This may not be the case in general and these terms may need to be considered when modelling more complex magnetic field configurations.

\bibliographystyle{jpp}

\bibliography{trm}

\makeatletter
\def\fps@table{h}
\def\fps@figure{h}
\makeatother

\end{document}